\documentclass[conference]{IEEEtran}
\IEEEoverridecommandlockouts
\usepackage{cite}
\usepackage{amsmath,amssymb,amsfonts}
\usepackage{algorithmic}
\usepackage{graphicx}
\usepackage{textcomp}
\usepackage{xcolor}

\usepackage{algorithm} 
\usepackage{algorithmic}
\usepackage[font=footnotesize]{subfig}
\usepackage{stfloats}
\usepackage{tcolorbox}
\usepackage{comment}

\def\BibTeX{{\rm B\kern-.05em{\sc i\kern-.025em b}\kern-.08em
    T\kern-.1667em\lower.7ex\hbox{E}\kern-.125emX}}
\begin{document}

\title{Reinforcement Learning Based Neighbour Selection for VANET with Adaptive Trust Management\\
}

\author{\IEEEauthorblockN{Orvila Sarker}
\IEEEauthorblockA{\textit{University of Adelaide} \\
\textit{Cyber Security Cooperative Research Centre}\\
orvila.sarker@adelaide.edu.au}
\and
\IEEEauthorblockN{Hong Shen}
\IEEEauthorblockA{\textit{University of Adelaide} \\
hong.shen@adelaide.edu.au}
\and
\IEEEauthorblockN{M. Ali Babar}
\IEEEauthorblockA{\textit{University of Adelaide} \\
\textit{Cyber Security Cooperative Research Centre}\\
ali.babar@adelaide.edu.au}

}

\maketitle

\begin{abstract}
Successful information propagation from source to destination in Vehicular Adhoc Network (VANET) can be hampered by the presence of neighbouring attacker nodes causing unwanted packet dropping. Potential attackers change their behaviour over time and remain undetected due to the adhoc nature of VANET. Capturing the dynamic attacker behaviour and updating the corresponding neighbourhood information without compromising the quality of service requirements is an ongoing challenge. This work proposes a Reinforcement Learning (RL) based neighbour selection framework for VANET with an adaptive trust management system to capture the behavioural changes of potential attackers and to dynamically update the neighbourhood information. In contrast to existing works, we consider \emph{trust} and \emph{link-life time} in unison as neighbour selection criteria to achieve trustworthy communication. Our adaptive trust model takes into account the \emph{social relationship}, \emph{time} and \emph{confidence} in trust observation to avoid four types of attackers. To update the neighbourhood information, our framework sets the learning rate of the RL agent according to the velocities of the neighbour nodes to improve the model's adaptability to network topology changes. Results demonstrate that our method can take less number of hops to the destination for large network sizes while can response is up to 54\% faster compared to a baseline method. Also, the proposed model can outperform the other baseline method by reducing the packet dropping rate up to 57\% caused by the attacker. 
\end{abstract}

\begin{IEEEkeywords}
VANET, Reinforcement Learning, Blackhole attack, Trust Management System
\end{IEEEkeywords}
\section{Introduction}
Vehicular Adhoc networks (VANETs) have been showing a burgeoning potential in facilitating safety critical applications such as emergency warnings, lane change assistance for developing future smart transportation systems \cite{sumayya2015vanet,wahid2019holistic}.  Large scale deployment of aforementioned applications are based on the assumption that each vehicle will collect, calculate and disseminate the information with other vehicles correctly. Decision making dependency on the perception of information received from other vehicles make VANET prone to wrong decisions and a wrong decision can lead to fatality \cite{javed2016interrelation}. 



The above discussion implies that decision making about intermediate nodes is critical. An inappropriate intermediate node selection can incur unnecessary packet dropping, high transmission delay or unwanted packet loss \cite{malhi2020security}. Trust based method is regarded as a widely adopted method in secure decision making about intermediate nodes in VANET \cite{hussain2020trust}. Trust is a means to quantify the quality of the information received from other neighbours in the network to evaluate the authenticity of the received information. However, trust calculation, establishment, maintenance and updating in VANET are challenging due to the change in node behaviour, network topology and absence of a centralised component for network monitoring \cite{mehdi2017game}.

Recently, Reinforcement Learning (RL) based methods for intermediate node selection are being adopted in dynamic wireless networks \cite{nazib2021reinforcement}. RL based methods have drawn significant attention to the networking community for several reasons, such as the inherent intelligence they bring into the system, their capability to cope with dynamic operating environment and suitability to solve network optimization problems in a distributed manner \cite{yau2012reinforcement}. Existing studies on RL based intermediate node selection can be broadly categorised into two main categories. One category of works (e.g., \cite{boyan1994packet,kumar1999confidence, wu2018reinforcement}) has considered the existence of network adversaries in the network. For instance, existing On-policy Monte Carlo (ONMC) based routing \cite{usaha2006identifying} intended to deal with network adversaries by prioritising highly secure neighbours for routing to reduce packet dropping by attackers. In another category of work (e.g., \cite{maneenil2005preventing, usaha2006identifying, zhang2018machine, zhang2018deep}), the design goal is to choose a neighbour that helps optimize one or more network performance metrics (e.g., packet delivery time \cite{boyan1994packet,kumar1999confidence}). 

\subsection{Motivation of this Study}\label{Challenges}

\textbullet\ Entities in VANET may change their behaviour over time. At time $t$, a node may forward the incoming packets correctly and at time $(t+1)$, it might drop all the received packets intentionally (e.g., Grayhole attack \cite{su2011prevention}). The main motivation of the attacker behind this is to gain benefit from the network \cite{soleymani2021security}. A trust management system should be effective enough to adaptively update the trust values of the intermediate nodes based on their dynamic behavioural pattern \cite{hussain2020trust}. 

\textbullet\ Safety critical applications are delay and latency sensitive \cite{hussain2019integration}, therefore it is important to optimise the QoS requirements of the network such as response time and the number of hop counts required to reach the destination. RL based routing is efficient than conventional non-adaptive algorithms (e.g., shortest path routing) for achieving improved Quality of Service (QoS) in dynamic networks \cite{nazib2021reinforcement}.
In RL based intermediate node selection techniques, the entire network is considered as the environment and each network node acts as an RL agent (e.g., \cite{boyan1994packet,kumar1999confidence,maneenil2005preventing, usaha2006identifying, zhang2018machine, zhang2018deep,wu2018reinforcement}). The neighbour selection has been performed based on some important security or QoS requirements of neighbours. As an example, in Q-routing \cite{boyan1994packet} and CQ-routing \cite{kumar1999confidence}, a node calculates, in the form of Q-value, the delivery time its neighbour requires to deliver packets to the destination. Q-values are stored in a Q-table and shared among other nodes. Finally, a node selects a neighbour for which the estimated Q-value is minimum (i.e., the neighbour which needs the least amount of time to deliver). Considering one or more QoS requirements of the network to prioritise intermediate nodes can improve network performance, but cannot guarantee security when attackers are present. For example, an attacker can share a wrong Q-value with its neighbour to increase its chances of being part of a route \cite{nazib2021reinforcement}. On the other hand, prioritising intermediate nodes based on security requirements can improve robustness against attacks \cite{zhang2018machine,zhang2018deep} but cannot guarantee QoS \cite{nazib2021reinforcement}. This necessitates the development of a neighbour selection technique that considers security while not compromising the QoS requirements of the network.

\textbullet\ A fast update of neighbourhood information is needed to capture the real-time traffic information on roads. For dynamic networks like VANET, updating neighbourhood information can be very challenging. Because of the distributed control property, a global view of the network is not possible \cite{zhou2020distributed}. Hence, the decision making of the node (agent) is entirely based on the local information available \cite{kumar1999confidence}. Learning rate plays vital role in updating the neighbourhood information  in RL based routing. The adjustable learning rate used in CQ-routing \cite{kumar1999confidence} causes excessive memory overhead when the network size gets larger. ARPRL \cite{wu2018reinforcement} introduced a variable learning rate considering the relative velocity difference of neighbours. However, in this method used the hello packet reception ratio to calculate the reward. Unfortunately, this can benefit the network adversaries (if any) who intentionally send frequent hello packets.
\subsection{Contribution of this Study}
Our main contributions are: \textbullet\ Development of an adaptive trust model that calculates the trust value of a node by considering social relationship, time of the observation and confidence in observation. Our adaptive trust model can capture the dynamic node behaviour and minimise the malicious effects of four different types of attackers. \textbullet\ Design of a Q-learning based framework that utilises both trust and link-life time for intermediate node selection in vehicular adhoc network. Incorporating trust helps to reduce the number of times an attacker is selected as an intermediate node and the number of packets dropped compared to a baseline method. Link-life time helps in selecting a high stable node. 
Incorporating security measures (trust) does not compromise the QoS requirements of the network. Our proposed framework still achieves good QoS performance compared to a baseline method.

Table \ref{comparison table} shows the comparison of the proposed framework with some of the related existing works.

\begin{table}[t]
\centering
\caption{Model comparison of proposed framework with existing works}
\label{comparison table}
\resizebox{!}{2cm}{
\begin{tabular}{lllllll}
\hline
Characteristics                                                   & CQ-routing  & \cite{zhang2018deep,zhang2018machine} & ONMC & Q-routing & ARPRL  & Proposed \\ \hline
Architecture                                                      & Distributed & Centralised & Distributed & Centralised & Distributed   & Distributed \\ \hline
Trust Model                                                       & - & Direct & Direct & - & - &   \begin{tabular}[c]{@{}l@{}}Direct, indirect\end{tabular}  \\ \hline
\begin{tabular}[c]{@{}l@{}}Trust\\calculation metric\end{tabular}       & - & PDR & PDR & - & - &   \begin{tabular}[c]{@{}l@{}}PDR, time,\\confidence \end{tabular} \\ \hline
\begin{tabular}[c]{@{}l@{}}Trust value\end{tabular} & - & - & Fixed & - & - &  Dynamic\\ \hline
\begin{tabular}[c]{@{}l@{}}Threat \\ model\end{tabular}         & - & \begin{tabular}[c]{@{}l@{}}Packet\\ disruption\\(Blackhole)\end{tabular} & \begin{tabular}[c]{@{}l@{}}Packet\\ disruption\\(Blackhole)\end{tabular} & - & - &   \begin{tabular}[c]{@{}l@{}}Packet disruption\\(Blackhole,\\ Grayhole),\\ Bad mouthing,\\ Ballot-stuffing \end{tabular} \\ \hline
\begin{tabular}[c]{@{}l@{}}Learning\\algorithm\end{tabular} & Q-learning & Deep Q-learning & \begin{tabular}[c]{@{}l@{}}On-policy \\monte carlo\end{tabular} & Q-learning & Q-learning&  Q-learning\\ \hline
\begin{tabular}[c]{@{}l@{}}Learning rate\end{tabular} & Adjustable & Fixed & Fixed & Fixed & Adjustable &  Adjustable\\ \hline
\begin{tabular}[c]{@{}l@{}}Learning\\ agent\end{tabular}          & \begin{tabular}[c]{@{}l@{}}Network\\ nodes\end{tabular} & \begin{tabular}[c]{@{}l@{}}SDN\\controller\end{tabular} & 
\begin{tabular}[c]{@{}l@{}}Network\\ nodes\end{tabular} & \begin{tabular}[c]{@{}l@{}}Network\\ nodes\end{tabular} & \begin{tabular}[c]{@{}l@{}}Network\\ nodes\end{tabular} & \begin{tabular}[c]{@{}l@{}}Network\\ nodes\end{tabular} \\ \hline
\begin{tabular}[c]{@{}l@{}}Reward\\ calculation\end{tabular}        & \begin{tabular}[c]{@{}l@{}}Data delivery\\ time\end{tabular} & \begin{tabular}[c]{@{}l@{}}Trust,\\ position\end{tabular} & Reputation & \begin{tabular}[c]{@{}l@{}}Data delivery\\ time\end{tabular} & \begin{tabular}[c]{@{}l@{}}Link-life,\\HMRR\end{tabular}  &   \begin{tabular}[c]{@{}l@{}}Trust,\\link-life\end{tabular}   \\ \hline
\begin{tabular}[c]{@{}l@{}}Mobility\\ awareness\end{tabular}      & No & Yes & No & No & Yes  & Yes  \\ \hline

\end{tabular}
}
\end{table}
\section{Problem Formulation in Q-learning Setting}
Our proposed framework employs a model free RL approach namely Q-learning (\cite{littman1993distributed}) to accomplish the neighbour selection task. In Q-learning the agent in each step observes the next state $s^{'}$ and calculates the expected maximum reward $r^{'}$ for the available set of actions $a^{'}$ in $s^{'}$ to update the Q-value of the corresponding action in the current state $Q(s,a)$ using the following equation:

\begin{equation}
\small
    \underbrace{Q(s,a)} _\text{updated value} \leftarrow  \underbrace{Q(s,a)}_\text{old value} + \lambda [r^{'} + \gamma \underbrace{max Q(s^{'}, a^{'})} _\text{maximum expected value} - \underbrace{Q(s,a)} _ \text{old value}]
\end{equation}

Here, $0< \lambda <1$ and $0<\gamma <1$ are the learning rate and discount factor respectively. Each node will asses the security status and link stability of its one-hop neighbours before routing packets to them. The security status of a neighbour is evaluated based on its trust value. In this work, we consider two types of packet dropping attacks namely Blackhole and Grayhole attacks and two types of trust management attacks namely bad mouthing and Ballot-stuffing attack. Link stability is estimated by calculating the link-life time of a neighbour. Link-life time is defined as the time a neighbour stays connected in direct communication before moving out of range. Selecting a neighbour with a longer link-life time saves both the time that would take to relaunch a new route and the amount of network resource required \cite{nabil2019predicting}.

Every node in the network will generate a preliminary set of one-hop neighbours by evaluating their trust value and store in a trust table. A Q-table consisting of Q-values of trustworthy neighbours will be formed from the trust table. In this work, the Q-value of a neighbour is an estimate of its link-life time for a particular destination. Larger link-life time results in a large Q-value. To route a packet, a neighbour having a high Q-value is selected.

VANET is a highly dynamic network where nodes move at high velocities causing a frequent change in the network topology. Therefore, the estimated Q-value of a neighbour becomes obsolete when it goes out of range due to mobility. Hence, an update of the Q-table is required to keep track of the currently available neighbours \cite{kumar1999confidence}. We calculate the learning rate of our Q-learning based routing model by using the velocity of the mobile nodes in the network. This adjustable learning rate helps adapt to any changes in network topology. 
\section{The Proposed Framework}
\subsection{System Model}
A VANET can be modelled as an undirected graph $G = \{ V \times E \}$ defined by a finite set of vertices $V=\{n_1,n_2,n_3...n_k \}$ where $n_i$ is a network node/vehicle (mobile or stationary) and a finite set of edges $E=\{e_1,e_2,e_3...e_m \}$. An edge $e_{ij}$ between node $n_i$ and node $n_j$ is defined by the following equation:
\begin{equation}
\small
    e_{ij} = L_{ij}
\end{equation}
Here, $L_{ij}$ is the link-life time i.e., the remaining time node $n_{i}$ will stay connected to $n_{j}$. Let, $T_{ij}$ = $\{n_{i}:n_{j}, Task\}$ be the trust level of node $n_{i}$ on node $n_{j}$ for performing the assigned \emph{Task}. In this work, the \emph{Task} is to forward the network packets correctly by following the protocol rules. Q-value at $n_{i}$ about sending a packet to the destination $D$ through neighbour $n_{j}$ is:
\begin{equation}
\small
\begin{aligned}
    Q_{n_i}(D,n_j) \leftarrow (1-\lambda_{n_{i}n_{j}}) Q_{n_i}(D,n_j) 
     + \\ \lambda_{n_{i}n_{j}} [R_{n_{i}n_{j}} +  {\max_{n_{k}\in N{_{T}(n_{j})}} Q_{n_{j}}(D, n_{k})}]
\end{aligned}
\end{equation}

where $\lambda_{n_{i}n_{j}}$ and $R_{n_{i}n_{j}}$ are the learning rate and constrained maximum reward respectively. Node $n_{k}$ is a one-hop neighbour of node $n_{j}$. Each node $n_{i}$ in the network maintains a Q-table consisting of Q-values $Q_{n_i}(D,n_j)$ that is an estimation of link-life time of trusted neighbour $n_{j}$ for destination $D$. Each network packet can be considered as an agent and change of states take place when packets move from one node to another. Action space consists of trusted one-hop neighbours. Routing decision is made based on a two-step evaluation of each neighbour node. At first, node $n_{i}$ generates a trusted neighbour set (action set) $N_{T}(n_{i})$ from all the one-hop neighbours $N(n_{i})$. Finally, from that trusted set, $n_i$ will calculate the expected maximum reward $R_{n_{i}n_{j}}$ $(R_{n_{i}n_{j}} = L_{ij})$ to choose an one-hop neighbour $n_j$ as an action if the Q-value $Q_{n_i}(D,n_j)$ for $n_j$ is maximum compared to other one hop neighbours $N(n_i)$ where $n_{j} \in N(n_i)$.
The calculation method of trust $T_{ij}$ and link-life time $L_{ij}$ in our proposed framework are described in Section \ref{Neighbourhood Information Calculation}.
\subsection{Attack Model}
\label{attack model}
In this work, we consider the four types of malicious activities by the attackers that can potentially compromise the security of the system: \textbullet\ Blackhole and Grayhole attack:
A Blackhole attack is a class of packet disruption attack where the attacker misleads legitimate nodes by sending them large sequence numbers and small hop count numbers to be part of a source-destination route. When the attacker becomes able to be part of an active route, it captures the packet from the legitimate nodes and drops it. On the other hand, a Grayhole attacker (alternately on-off attack) drops selective packets and forwards the rest of them \cite{su2011prevention}. \textbullet\ Bad mouthing attack:
In this type of attack, an adversary spreads unfair trust ratings about non-malicious nodes with the intention to decrease their overall trust values in the network \cite{kudva2021scalable}. \textbullet\ Ballot-stuffing attack: In this type of attack, a malicious entity propagates exaggerated trust value of poorly performed nodes in order to make the trust management system predict them as highly trusted \cite{shabut2014recommendation}.

\subsection{Neighbourhood Information Update} The quality of the route depends on how closely the Q-values can reflect the current state of the network. Node velocities are used to calculate our adjustable learning rate. A node $n_i$ moving at a velocity $V_{i}$ will calculate the learning rate $\lambda_{{n_i}n_j}$ of its neighbour $n_j$ that moving at a velocity $V_{j}$ as follows:
\begin{equation}
    \small
    {\lambda_{{n_i}n_j}} =  \frac{|(V_{i}-V_{j})|}{(V_{max}-V_{min})},  if  {|(V_{i}-V_{j})|} > V_{th}
\end{equation}
Here, $V_{max}$, $V_{min}$ and $V_{th}$ are maximum velocity, minimum velocity, velocity threshold respectively. Equation (4) indicates that the learning rate will dynamically increase when the speed difference between two nodes is high. And when the speed difference is below a given velocity threshold, the routing policy will be updated according to a fixed value. Each node reactively updates its trust table and Q-table through periodically exchanged HELLO packets and RREQ/RREP messages. When an available route becomes invalid (due to HELLO packet time-out), the Q-value of a neighbour is reset to $0$.




\subsection {Neighbourhood Information Calculation} \label{Neighbourhood Information Calculation}
In our proposed trust model, direct trust is calculated by a node itself using Bayes theorem \cite{pearl2014probabilistic} and recommendations are gathered from one-hop neighbours to compute the indirect trust using Yager's rule \cite{yager1987dempster}. Yagers' rule is a modified version of well-known Dempster Shafer Theory \cite{shafer1992dempster}. Bayes theorem offers the advantage of incorporating the previous information about a nodes' trust and helps to form a prior distribution for future trust calculation \cite{wei2014security}. Moreover, it provides good defense against Bad mouthing and Ballot-stuffing attack \cite{hoffman2009survey}. Yager's rule offers improved performance when multiple information are completely conflicting \cite{zadeh1979validity}.
\subsubsection{Direct Trust Calculation} 
\label{Direct Trust Calculation}
Let, $P = \{n_{i}:n_{j}, Task\}$ be the probability that node $n_{j}$ will perform the task assigned by node $n_{i}$, and $Z$ be a random variable that defines the degree of belief where $0\leq z \leq 1$. If a node receives a total $N$ number of packets and successfully forwards $l$ packets then the posterior probability or belief function can be calculated using Bayes theorem using the equation below:

\begin{equation}
\small  
    f(z, N|l) = \frac{P(l|z,N) f(z,N)}{\int P(l|z, N) f(z,N) dz}
\end{equation}

The likelihood function $P(l|z,N)$ is the probability of forwarding k packets out of the total N received packets. Assume that $P(l|z,N)$ follows Binomial distribution:

\begin{equation}
  \small 
    P(l|z,N) = \frac{N!}{l!(N-l)!} z^l (1-z)^{N-l}
\end{equation}
Assuming Binomial distribution is reasonable in this case as the observation (trust value of a neighbour) is either success (trustworthy) or failure (malicious). Also, each observation is fixed and independent.
The prior probability $f(z,N)$ can be calculated as Beta distribution on parameters $\alpha$ and $\beta$:

\begin{equation}
\small    
    f(z|\alpha, \beta)= \frac{z^{\alpha -1}(1-z)^{\beta - 1}}{B(\alpha, \beta )}
\end{equation}

where $\alpha$ and $\beta$ are the number of successful and unsuccessful packets forwarded by the node respectively and $\alpha$ , $\beta$ $>$ 0.  The mean $\mu$ i.e., the expected value of a Beta distribution random variable Z on $\alpha$ and $\beta$ is:
\begin{equation}
\small    
    \mu = E(Z|\alpha, \beta) = \frac{\alpha}{\alpha+\beta}\\
 \end{equation}  

   \begin{equation}
   \small
    \sigma = Var(Z| \alpha, \beta) = \frac{\alpha \beta}{{(\alpha + \beta)}^2 (\alpha + \beta +1)} 
 \end{equation}
\begin{table}[]
\setlength{\tabcolsep}{1pt}
\caption{Common notations used in this study}
\label{tab:my-table}
\resizebox{!}{1.45cm}{
\begin{tabular}{ll|ll}
\hline
Symbol & Description & Symbol & Description \\ \hline
    $N$  & Total number of nodes  & $S$  & Source/Forwarding node           \\ \hline
    $T_{th}$     &  Trust threshold           &   $D$      &   Destination node          \\ \hline
      $\alpha_{q}$   &  \begin{tabular}[c]{@{}l@{}}Number of packets send at time $t_{q}$ \end{tabular}            &   $L_{ij}$       & Link-life time between node $n_{i}$ and node $n_{j}$             \\ \hline
        $\beta_{q}$   &    Number of packets dropped at time $t_{q}$         &        $R_{n_{i}n_{j}}$  & \begin{tabular}[c]{@{}l@{}}Received reward of node $n_{i}$ for selecting\\ node $n_{j}$\end{tabular}             \\ \hline
         $N_{T}(n_{i})$&     Trusted neighbour list of node $n_{i}$        & $\lambda_{n_{i}n_{j}}$         & \begin{tabular}[c]{@{}l@{}}Learning rate for neighbour $n_{j}$ calculated by\\ node $n_{i}$ \end{tabular}           \\ \hline
        $T_{ij}$ &   Direct trust of node $n_{i}$ on node $n_{j}$          &  $Q_{n_{i}}(D,n_{j})$        &  \begin{tabular}[c]{@{}l@{}} Q-value of neighbour $n_{j}$ calculated by node\\ $n_{i}$ for destination $D$ \end{tabular}            \\ \hline
       $G_{ij}$  &   Indirect trust of node $n_{i}$ on node $n_{j}$          &  $V_{i}$        &  Velocity of node $n_{i}$           \\ \hline
        $T_{ij}^t$ &   Total trust of node $n_{i}$ on node $n_{j}$           & $V_{th}$         & Velocity threshold            \\ \hline   
\end{tabular}
}
\end{table}
\subsubsection{Updating Direct Trust}
In order to take into consideration the full ignorance when there is no observation made about a node at the beginning, the prior probability $f(z,N)$ is assumed to be equal to $B(1,1)$. This means the initial trust value of each node before any evaluation is equal to 0.5. The trust values can be updated through continuous observations. After $q$ number of interactions, the expected value can be calculated as,
\begin{equation}
\small    
    E_{q}(Z) = \frac{\alpha _{q}}{\alpha _{q} + \beta_{q}}
\end{equation}
\subsubsection{Freshness in Direct Trust Estimation}
\label{Freshness in Direct Trust Estimation}
While calculating the trust value, it is useful to give the recent interactions/observations more weights to minimise the adverse effects of Grayhole attack  \cite{sun2006trust}. We introduce a diminishing factor to achieve this. Consider at time $t_{q-1}$, a node has successfully forwarded $\alpha_{q-1}$ packets. At time $t_q$, the number of packets forwarded by the node will be calculated by the following equation:
\begin{equation}
\small    
    \alpha_{q} = \alpha_{q-1}.(c)^{\Delta t}
\end{equation}
 where $\Delta t= {(t_{q}-t_{q-1})}$ is an integer and $0<c< 1$.
\subsubsection{Indirect Trust Calculation}
\label{Indirect Trust Calculation}
Indirect trust is calculated using Yagers' rule. A frame of discernment $\Psi$, consisting of a mutually exclusive and exhaustive set of propositions are considered. A  subset of $\Psi$ including itself and $\emptyset$ is termed as focal elements where $2^{\Psi}$ = $\{X_1,\Psi, X_n\}$ where, $X$ is any  hypothesis or proposition.
The probability of a focal set $X$ is a function m: $2^{\Psi}$ $\rightarrow$ $[0, 1]$, m satisfies the following conditions: $m(\emptyset) = 0$, and $\sum_{X_i \subseteq \Psi} m(X_i)=1$.

\begin{figure}
 \centering
   \includegraphics[trim=220 140 220 575,clip, scale = 0.6]{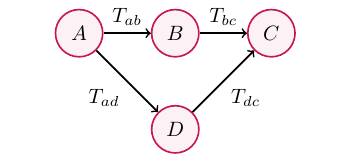}
  \caption{Trust establishment among network nodes}
  \label{Trust Establishment among Network Nodes}
\end{figure}

In our proposed trust model, the security status of a node $n_i$ can be either normal or malicious similar as applied in (\cite{chen2005dempster}). As the security states of vehicles are mutually exclusive, the frame of discernment is defined as $\Psi$ = $\{{\tau},{\tau ^{'}} \}$ where $\tau$ = $n_i$ is normal node, ${\tau ^{'}}$ = $n_i$ is malicious node. Three focal elements of $\Psi$ are given below:
\[
    \text{Hypotheses}
    \left\{
    \begin{array}{lr}
      H = \tau & \text{if } n_{i} \text { is normal} \\
      H^{'} = \tau^{'} & \text{if } n_{i} \text { is malicious} \\
      U = \Psi & \text{if } n_{i} \text { either normal/malicious}
    \end{array}
    \right\} 
  \]
Consider in Figure \ref{Trust Establishment among Network Nodes}, node $A$ wants to send message to node $C$ which is not directly connected to $A$ and suppose $A$ has no information about $C$ in its current trust table. In such cases, node $A$ seeks for recommendation to its one hop neighbours which are node $B$ and node $D$. Recommendations from $B$ and $D$ about $C$ can be any of the following: \textbullet\  \emph{\textbf{Case 1:}} $B$ recommends $C$ as trustworthy and $D$ recommends $C$ as untrustworthy, \textbullet\  \emph{\textbf{Case 2:}} $B$ recommends $C$ as untrustworthy and $D$ recommends $C$ as trustworthy, \textbullet\  \emph{\textbf{Case 3:}} Both $B$ and $D$ recommends $C$ as trustworthy, \textbullet\  \emph{\textbf{Case 4:}} Both $B$ and $D$ recommends $C$ as untrustworthy.

For case 1, $B$ supports hypothesis $H$ and $D$ supports hypothesis $H^{'}$, for case 2, $B$ supports hypothesis $H^{'}$ and $D$ supports hypothesis $H$ and so on. So, there might be uncertainties about the recommendation node $A$ receives from its direct neighbours $B$ and $D$ which depends upon the security status of $B$ and $D$. This is because it might be the case that either $B$ or $D$ is providing false recommendation about $C$ intentionally. In such cases, uncertainties can be reduced by considering belief and plausibility (\cite{chen2005dempster}) of the observation. 
 \subsubsection {Trust as Belief and Plausibility}
An observer can combine independent pieces of evidences to obtain the degree of belief about a hypothesis from the corresponding subjective probabilities. For any subset Y of $\Psi$, the belief function is defined as:
\begin{equation}
    bel(Y) = \sum_{X_j \subseteq Y} m(X_j) =1
\end{equation}
It can be interpreted that the basic probability value $m(X)$ is the portion of the total belief assigned to hypothesis $X$ out of $j$ propositions, which reflects the evidence's strength of support. Weight of evidence that does not refute $Y$ maps each hypothesis Y to a value $Pls(Y)$ between 0 and 1.

\begin{equation}
\small   
    Pls(Y) = \sum_{j:X_j \cap Y \neq \emptyset}m(X_j)
\end{equation}
Here, Pls(Y) denotes plausibility and $Pls(Y) = 1 - bel(Y^{'})$

Consider $m_1(X)$ and $m_2(X)$ are the subjective probabilities from two independent observers (in the same frame of discernment). Dempster rule to combine the probabilities is $m(Y)$ = $m_1(Y)$ $\oplus$ $m_2(Y)$. In our proposed trust model, the direct trust value is assigned as the basic probability value of a node. In our example shown in Figure \ref{Trust Establishment among Network Nodes}, the direct trust value $T_{ab}$ of node A has on node B is calculated by Bayesian inference that we discussed in the Section \ref{Direct Trust Calculation}. If node B claims that C is a normal node, the basic probability assignment to each hypothesis will be $m_B(H) =T_{ab}$, $m_B(H^{'}) =0$, and $m_B(U) =1- T_{ab}$. Alternatively, if node B claims C as malicious, its basic probability to each hypothesis will be, $m_B(H) =0$, $m_B(H^{'}) =T_{ab}$, and $m_B(U) =1- T_{ab}$. Table \ref{table_mass_assignment} shows the mass probabilities for each hypothesis.
Therefore, the belief function of each focal set in $\Psi$ can be calculated as:
\begin{equation}
\small
\begin{aligned}[b]
    bel_{B}(H) = m_{B}(H),
    bel_{B}(H^{'}) = m_{B}(H^{'}),\\
    bel_{B}(U) = m_{B}(H)+m_{B}(H^{'})+m_{B}(U)
\end{aligned}
\end{equation}
\begin{table}
\scriptsize
\renewcommand{\arraystretch}{1} 
\caption{Mass function assignment}
\label{table_mass_assignment}
\centering
\begin{tabular}{c c c c}
\hline
Hypotheses & $m_{B}(H)$ & $m_{B}(H^{'})$ & $m_{B}(U)$ \\
\hline
$H =\{\tau\}$ & $T_{ab}$ & $0$ & $1-T_{ab}$ \\
$H^{'}=\{\tau^{'}\}$ & $0$ & $T_{ab}$ & $1-T_{ab}$ \\
\hline
\end{tabular}
\end{table}
\subsubsection{Combining Multiple Recommendations} 


If $bel_{1}(Y)$ and $bel_{2}(Y)$ are two belief functions on the proposition $Y$ over the same frame of discernment $\Psi$, then according to Yagers' rule, the total belief $bel(Y)$ on proposition Y can be calculated by the following equation:
\begin{equation}
bel(Y) = {\sum_{i,j: {X_{i}\cap X_{j}} = Y} {m_{1}}(X_{i})m_{2}(X_{j})},  for Y \neq \emptyset, Z
\end{equation}

\begin{equation}
  bel(Z) = {\sum_{i,j: {X_{i}\cap X_{j}} = Z} {m_{1}}(X_{i})m_{2}(X_{j}}) + log(1 - k)
\end{equation}

Here, $k = {\sum_{i,j: {X_{i}\cap X_{j}} \neq \emptyset} {m_{1}}(X_{i}){m_{2}}(X_{j})}$ measures the amount of disagreement in collected indirect trust opinions.

In the example shown in Figure \ref{Trust Establishment among Network Nodes}, node $A$ calculate the combined belief of node $B$ and node $D$ on hypothesis $H$ and $H^{'}$ by the following equations:
\begin{equation}
\begin{aligned}
 \scriptsize 
    m_{B}(H) \oplus m_{D}(H) = \frac{1}{M}[m_{B}(H)m_{D}(H)+\\ m_{B}(H)m_{D}(U)+ m_{B}(U)m_{D}(H)]
\end{aligned}
\end{equation}
 \begin{equation}
 \small
 \begin{aligned}    
     m_{B}(H^{'}) \oplus m_{D}(H^{'}) =
    \frac{1}{M}[m_{B}(H)+m_{D}(H^{'})+\\m_{B}(H)m_{D}(U)+m_{B}(U)m_{D}(H^{'})]
 \end{aligned}
 \end{equation}
\begin{equation}
  \small 
     m_{B}(U) \oplus m_{D}(U) =
    \frac{1}{M}m_{B}(U)m_{D}(U)
 \end{equation}
Here, $M = m_{B}(H)m_{D}(H) + m_{B}(H)m_{D}(U) + m_{B}(U)m_{D}(H) + m_{B}(H^{'})m_{D}(H^{'}) + m_{B}(H^{'})m_{D}(U) + m_{B}(U)m_{D}(H^{'}) + m_{B}(U)m_{D}(U)$.

\begin{algorithm}[t]
\scriptsize
\caption{Trusted neighbour list creation by node $n_{i}$}
\label{Trusted Neighbour List Creation by Node $n_{i}$}
\begin{algorithmic}[1]
\STATE \textbf{Input:} $T_{th}$, $\alpha_{q}$, $\beta_{q}$, total number of nodes $N$
\STATE \textbf{Output:} Trusted neighbour list $N_{T}(n_{i})$ of node $n_{i}$
\FOR{each time stamp $t_q$} 
    \STATE Calculate time difference, $\Delta t = {t_q} - {t_{q-1}}$
   \FOR{each neighbour $n_{j}$ in $N$} 
    \STATE Calculate confidence factor $\rho = \frac{12 \alpha_{q} \beta_{q}}{(\alpha_{q} + \beta_{q})^2 (\alpha_{q} + \beta_{q} + 1)}$
    \STATE Calculate direct trust ${T_{ij}} = \frac{\alpha_{q}}{\alpha_{q} + \beta_{q}}$
    \STATE Calculate indirect trust $G_{ij}$ using Equation (16)
    \STATE Calculate total trust $T^{t}_{ij} = \rho T_{ij} + (1 - \rho) G_{ij}$
    \IF {$T^{t}_{ij}$ $>$ $T_{th}$}
    \STATE Add neighbour $n_{j}$ to the trusted list $N_{T}(i)$
    \STATE Update $\alpha_{q}$: $\alpha_{q}  \leftarrow \alpha_{q}.(c)^{\Delta t}$
    \STATE Update $\beta_{q}$: $\beta_{q}  \leftarrow P - \alpha_{q} $
    \ELSE
    \STATE Ignore neighbour $n_{j}$
    \ENDIF 
   \ENDFOR
\ENDFOR 
\end{algorithmic}
\end{algorithm}
\subsubsection{Total Trust Calculation}
A node $n_{i}$ will calculate the total trust value $T^{t}_{ij}$ of another node $n_{j}$ by combining the direct observation trust $T_{ij}$ and indirect observation trust $G_{ij}$ by using the following equation.
\begin{equation}
    T^{t}_{ij} = \rho T_{ij}+G_{ij}(1-\rho)
\end{equation}
where $\rho$ is the weighting coefficient we termed as confidence factor. The indirect trust value $G_{ij}$ is calculated by fusing multiple belief functions from one hop neighbours using Equation (16). It is important to provide more weight to the direct observation in the overall trust calculation to minimize the effect of wrong recommendations coming from malicious nodes. If $T^{t}_{ij}$ is greater than a trust threshold $T_{th}$, a neighbour is considered as normal or legitimate. Algorithm \ref{Trusted Neighbour List Creation by Node $n_{i}$} demonstrates the overall trust calculation. 

\subsubsection{Calculation of Confidence Factor, $\rho$ }
If a neighbour successfully forwarded $\alpha$ number of packets out of received $(\alpha + \beta)$ number of packets, where $\beta$ is the number of packets dropped, confidence factor $\rho$ is calculated by Equation \ref{rho calculation}.  

\begin{equation}
\label{rho calculation}
    \rho = \frac{12 \alpha \beta}{(\alpha + \beta)^2 (\alpha + \beta +1)}
\end{equation}
A value of $\rho$ close to 1 indicates high confidence in observation. If the value of $\rho$ is equal to 0, a neighbours' total trust value is calculated only on the basis of direct observation (i.e., direct trust). On the other hand, if there is no direct trust relationship with a neighbour ($\rho = 0$), total trust is calculated from the recommendation (indirect trust) received from other neighbours'. 
\begin{algorithm}[t]
\scriptsize
\caption{Q-value calculation}
\label{Q-value Calculation}
\begin{algorithmic}[1]
\STATE \textbf{Denotation:} Initial state/source  =  $S$, goal state/destination =  $D$, current state/node = $v_c$, neighbour of $v_{c}$ = $n$, neighbour of $n$ = $n^{'}$
\STATE \textbf{Input:} Trusted neighbour set $N_{T}(v_{c})$, $V_{max}$, $V_{min}$, $T_{th}$, $V_{th}$, $Q(D,n)$
\STATE \textbf{Output:} Q value of neighbour $n$ calculated by source/forwarding vehicle $v_{c}$
\STATE \textbf{Initialize:} Q-table for all $n$
\STATE Available action set $A_{v_{c}} = N_{T}(v_{c})$
\FOR {each trusted neighbour $n \in A_{v_{c}}$}
\IF{$(V_{v_{c}} - V_{n}) \geq V_{th}$}
\STATE Calculate learning rate, $\lambda_{v_{c},n} = \frac{(V_{v_{c}} - V_{n})}{V_{max} - V_{min}}$
\ELSE
\STATE $\alpha_{v_{c},n} = 0$
\ENDIF
\STATE Calculate constrained reward $R_{v_{c},n}$ = $L_{v_{c},n}$
\STATE $Q(D,n) \leftarrow (1-\lambda_{v_{c},n})Q(D,n) + \lambda_{v_{c},n} [R_{v_{c},n} + max Q(D, n^{'})]$
\ENDFOR
\end{algorithmic}
\end{algorithm}

\subsubsection{Neighbour Link-life Time Calculation}
\label{Neighbour Link-life Time Calculation}
At any given time, the link-life time $L_{{ij}}$ between two nodes $n_i$ and $n_j$ is the time the vehicles remain directly connected. For neighbour link-life time calculation, we adopt the method discussed in \cite{nabil2019predicting}. We assume that each node can collect its location, speed and direction information from the GPS equipped with it. Also, each node sends information like location, speed, direction, vehicle id and current time to its directly connected neighbours through exchanging periodic beacon messages. The distance between two nodes on the ordinate axis is considered negligible compared to the transmission range $R$. 
Let us consider that the two nodes $n_{i}$ and $n_{j}$ are moving with velocities $V_{i}$ and $V_{j}$. Assume that, at time $t_{1}$ they are in the position $(x_{i},y_{i})$ and $(x_{j},y_{j})$, and at time $t_{2}$ the new positions of the two nodes are $({x_{i}^{'}},{y_{i}^{'}})$ and ${(x_{j}^{'}},{y_{j}^{'}})$, respectively. If node $n_{i}$ and $n_{j}$ are moving with a different velocities and they have a constant acceleration during direct communication, the link-life time $L_{ij}$ between them can be calculated by the following equation:
\begin{equation}
{L_{ij}} = 
\begin{cases}
    \frac{d^{'}-d}{V_{i}-V_{j}}, & \text{nodes moving in the same direction}\\
    \frac{d^{'}-d}{V_{i}+V_{j}},& \text{nodes moving in the opposite direction}\\
\end{cases} 
\end{equation}
Here $d$ and $d^{'}$ are the distances travel in the $x$ direction at time $t_{1}$ and $t_{2}$ respectively. 
Algorithm \ref{Q-value Calculation} demonstrates the steps involved in learning and updating the Q-value of a neighbour in proposed framework.

\begin{figure}[t!]
 \centering
   \includegraphics[width=.49\textwidth]{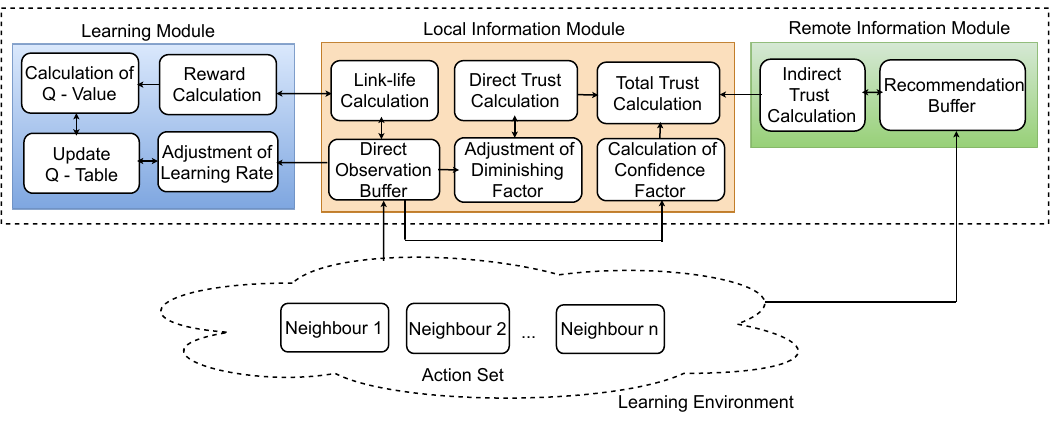}
  \caption{Main elements of proposed pramework}
  \label{High Level Overview of QSAR}
\end{figure}
\subsection{Main Components of Proposed Framework}
Figure \ref{High Level Overview of QSAR} displays a simplified block diagram of proposed framework. Proposed framework comprised of the following three main components:
\textbullet\ Local Information Module:
The observation buffer in the local information module of node $n_{i}$ keeps track of the link status and the number of packets received/forwarded by neighbour $n_{j}$. Based on the number of packets successfully forwarded by $n_{j}$, node $n_{i}$ adjusts the diminishing factor $c$ and confidence factor $\rho$ to calculate the direct trust $T_{ij}$ of $n_{j}$. Indirect trust values are extracted from the remote information module and combined with direct trust to calculate the total trust $T^{t}$.
\textbullet\ Remote Information Module:
The remote information module of a node $n_{i}$ collects trust recommendations about a neighbour $n_{j}$ experienced by other one-hop neighbours $N(n_{i})$ to calculate indirect trust $G_{ij}$.
\textbullet\ Learning Module:
The task of the learning module is to calculate, store and update the Q-values of trusted one-hop neighbours. Rewards for neighbours are obtained from the local information module in the form of link-life time. Link-life time is calculated only for the trusted neighbours. Link status of neighbours monitored by direct observation buffer is extracted to calculate the learning rate. The Q-table is of the size $(n_{r}*n_{r})$ where $\{n_{r}, \forall r \}$ denotes the benign nodes (trust values greater than $T_{th}$) on a given available route.
\section{Performance Evaluation}
To illustrate the performance of proposed framework, we perform simulations in ns-2 \cite{ns2} on Ubuntu 18.04 and Python 3.6 on Windows 10 64-bit operating system (Intel(R) Core(TM) i7-8700 CPU with 15.8GB memory). Due to the stochastic nature of the reinforcement learning algorithm, the data points in Figure \ref{Performance of QSAR in security context} and Figure \ref{comparison_between_QSAR_ARPRL} are plotted with 90\% confidence intervals (each result averaged over 100 randomly seeded executions). The main simulation parameters used in this work are tabulated in Table \ref{Simulation Setup}. The trust scores of each node obtained from ns-2 are used to create a trust matrix to simulate the proposed framework in Python. Similarly, node velocities are stored in a velocity matrix to calculate the variable learning rate $\lambda$. 
\begin{table}[!t]
\scriptsize
\renewcommand{\arraystretch}{1} 
\caption{Simulation setup}
\label{Simulation Setup}
\centering
\begin{tabular}{c c c c c c c c c c c}
\hline
Parameter & Symbol & Value\\
\hline
Network Topology & - & Random and Grid\\
Node Mobility & - & Static and Mobile\\
Node Velocities & $V_{i}, V_{j}$ & $(5 - 35)$ $ms^2$\\ 
Number of Nodes & $N$ & $4, 8, 16, 32, 64$\\
Maximum Velocity & $V_{max}$ & $45$ $ms^2$\\
Minimum Velocity & $V_{min}$ & $5$ $ms^2$\\
Initial Q-Value & $Q_{n_{i}}(D,{n_{j})}$ & $0$\\
\hline
\end{tabular}
\end{table}

\subsection{Modelling Trust Variation over Time}
\label{Modelling Trust Variation over Time}
\begin{figure}[!t]
\centering
\includegraphics[trim=15 310 635 5,clip, scale = 0.50]{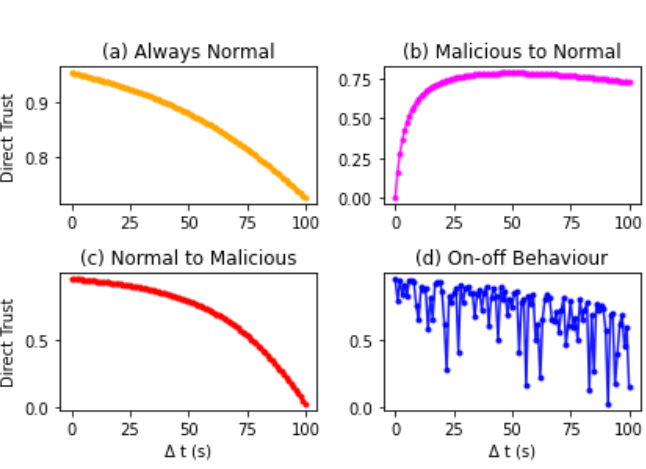}
\caption{Change of direct trust of a node over time}
\label{alpha_variations}
\end{figure}
Figure \ref{alpha_variations} shows how diminishing factor causes the change of direct trust of a node over time. We consider four cases to demonstrate the change of trust value of a randomly selected node based on its behavioural changes over time: \textbullet\  when a node always forwards packets correctly (Figure \ref{alpha_variations}a) \textbullet\ when a node starts behaving normal from being malicious (Figure \ref{alpha_variations}b) \textbullet\ when a node starts behaving malicious from being normal (Figure \ref{alpha_variations}c) \textbullet\ when a nodes' behaviour fluctuates (normal and malicious alternatively) over time (Figure \ref{alpha_variations}d).

Direct trust of a node decreases over time even if it always forwards packets correctly (Figure \ref{alpha_variations}a). This helps to capture the freshness in trust estimation. A nodes' direct trust increase over time when it turns into normal (Figure \ref{alpha_variations}b). This gives good nodes the opportunity to regain high trust if their trust has been decreased previously due to undesirable circumstances (e.g., environmental changes like bad wireless channel conditions). A nodes' trust tends to decrease if it changes its' behaviour from normal to malicious. On the other hand, a node's trust fluctuated time to time if it behaves normal and malicious alternatively (Figure \ref{alpha_variations}d). This change in trust helps to encourage bad or selfish nodes to be consistent in their normal behaviour.
\subsection{Accuracy of Trust Estimation}
We compare the adaptive trust model with a fixed trust approach where the values of $\rho$ are 0.2, 0.5 and 0.8 throughout the simulation time. In this comparison, we have not considered the effect of diminishing factors. This choice is justified by our intention to demonstrate our adaptive trust models' capability when time dependency is not considered. 

\begin{figure*}[t!]
\centering
\subfloat[Trust Value of a Highly Trusted Node]{\includegraphics[trim=3 325 650 4,clip, scale = 0.40]{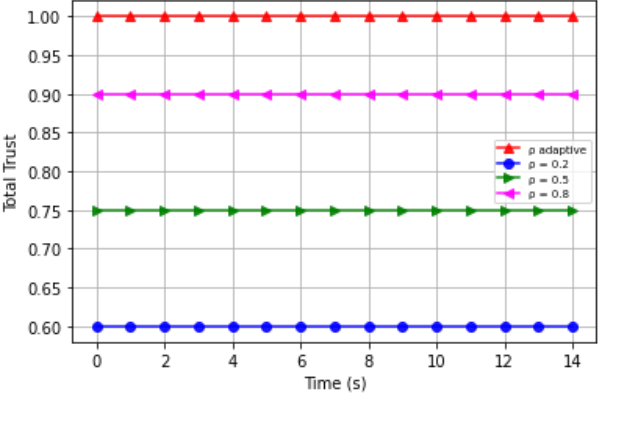}%
\label{highly_trusted_node}}
\hfil
\subfloat[Trust Value of a Blackhole Attacker]{\includegraphics[trim=3 325 650 4,clip, scale = 0.40]{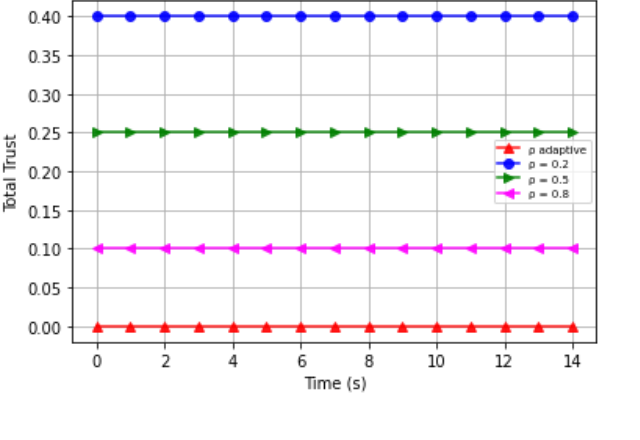}%
\label{maliciou_node_blackhole_attacker}}
\hfil
\subfloat[Trust Value of a Trusted Node]{\includegraphics[trim=3 325 635 0,clip, scale = 0.40]{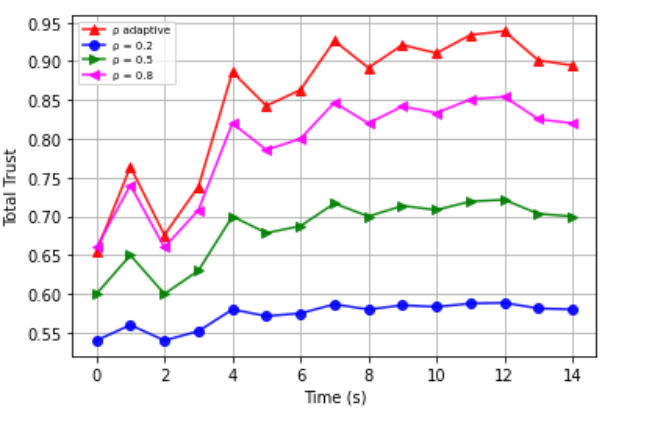}%
\label{trust_normal_node}}
\hfil
\subfloat[Trust Value of a Malicious Node]{\includegraphics[trim=3 325 635 0,clip, scale = 0.41]{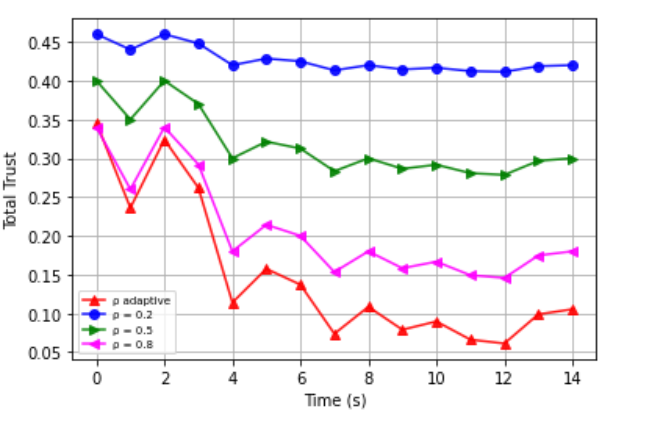}%
\label{trust_malicious_node}}
\caption{Trust variations of normal and malicious nodes}
\label{X}
\end{figure*}

For a highly trusted node who always sends packets correctly (Figure \ref{highly_trusted_node}), our adaptive trust model provides more accurate trust value of the node. On the other hand, a malicious node who always drops packet (Figure \ref{maliciou_node_blackhole_attacker}) receives 0 total trust from our trust model. Again when a node drops few packets unintentionally (e.g., due to bad wireless channel conditions or high mobility) but the number of forwarded packets are still greater that the number dropped, it deserved to get a higher trust ratings. Figure \ref{trust_normal_node}  shows our adaptive trust model provides relatively higher trust value for trusted node like this. Alternatively, our adaptive trust model generates lowest trust (Figure \ref{trust_malicious_node}) in cases when a malicious node forwards negligible amounts of packets compared to the number of packets it drops.



\begin{figure*}[t!]
\centering
\subfloat[Effect of fixed and dynamic trust status]{\includegraphics[trim=100 265 110 280,clip, scale = 0.3]{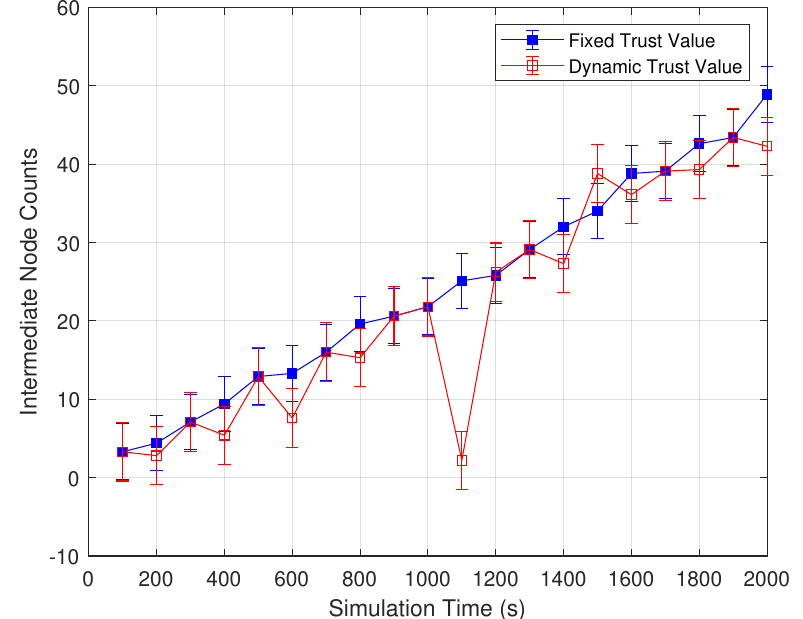}%
\label{trust}}
\hfil
\subfloat[Blackhole attacker as a forwarding node]{\includegraphics[trim=100 265 110 280,clip, scale = 0.3]{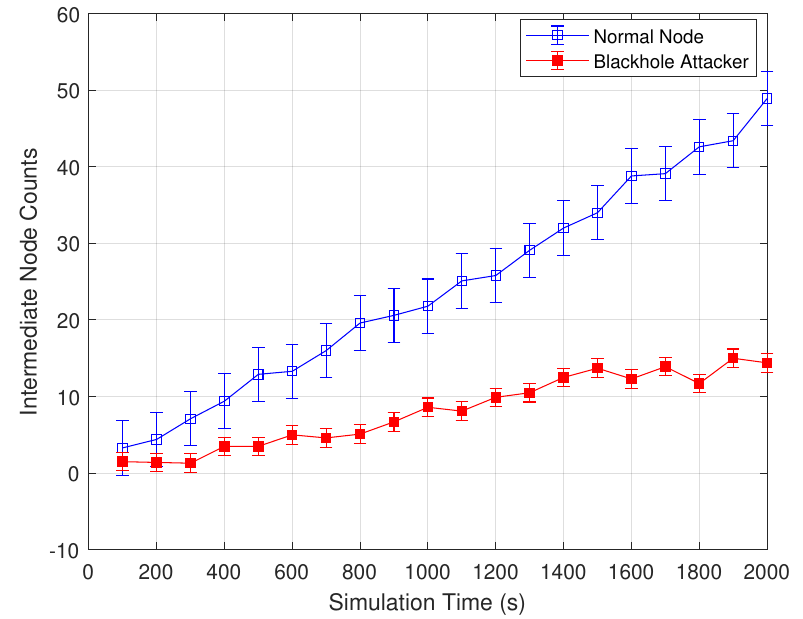}%
\label{blackhole}}
\hfil
\subfloat[Grayhole attacker as a forwarding node]{\includegraphics[trim=100 265 110 280,clip, scale = 0.3]{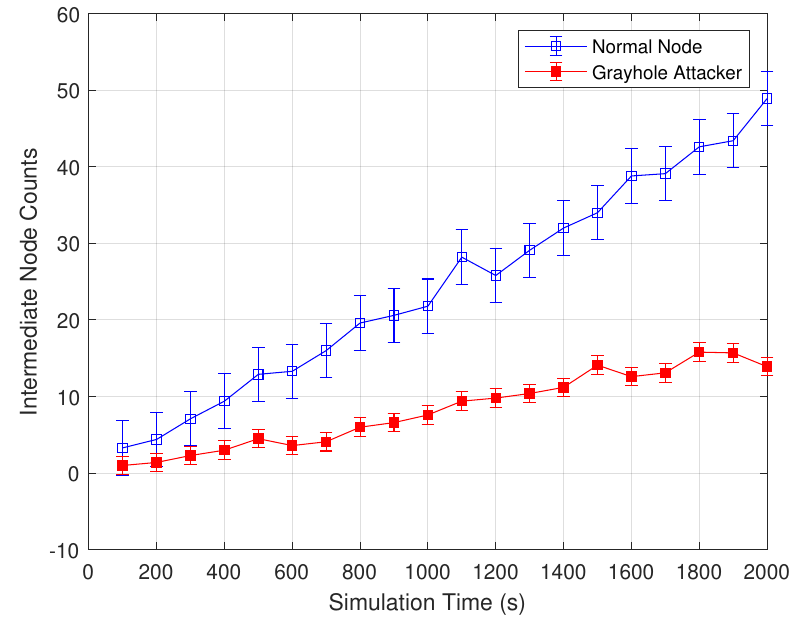}%
\label{grayhole}}
\caption{Performance of proposed framework in security context (total legitimate nodes = 63, attacker node = 1)}
\label{Performance of QSAR in security context}
\end{figure*}
\subsection{Effect of Fixed and Variable Trust}
In proposed framework, a node's trust dynamically changes over time according to the value of diminishing factor. It is expected that when a node's trust decreases, the number of times it served the role as a forwarding node will also be reduced. To demonstrate this effect, we introduce a metric \emph{intermediate node count} to calculate the number of times a node has been selected by a source or its neighbour nodes to reach a particular destination. The intermediate node count of a high stable intermediate neighbour node should be higher if link-life time is considered for neighbour selection. We randomly select a node in the network to show how its role as an intermediate node evolves with its trust status (Figure \ref{trust}). When the trustworthiness remains fixed, its chances of being a forwarder remains stable and monotonically increases with time. On the other hand, we observed that when we selectively decrease the trust value in between the simulation time, a node's role as a forwarder fluctuates. Intermediate node count is slightly higher when the trust value of the node is high and it decreases with a decrease in node trustworthiness.
\subsection{Robustness Against VANET Attacks}
A node's intermediate node count can be an important metric in the security context. The total intermediate node counts measure how many times a node was part of a route. In a routing model, intermediate node counts of an attacker should be as low as possible to minimise packet dropping. Therefore, we calculate the intermediate node counts of an attacker to demonstrate the resiliency of proposed framework against attacks. To illustrate the effect of the Blackhole/Grayhole attacker, we tested proposed framework on a network of 64 nodes where 63 of them are benign (normal) nodes and 1 is malicious (Blackhole/Grayhole). Figure \ref{blackhole} and Figure \ref{grayhole} show that the attacker's chances of being part of a route are reduced up to 70\% compared to a normal node. Moreover, even though a Grayhole attacker does not drop all the packets it receives, its count of intermediate nodes is as low as a Blackhole attacker. This is because proposed framework collects recommendation from other neighbours in the form of indirect trust which suppresses a Grayhole attacker's ability to act as benign and malicious alternatively (on-off). 
\subsection{Robustness against Trust Management Attacks}
\label{Robustness against Trust Management Attacks}
\begin{figure}[!t]
\centering
\subfloat[Total trust values of a normal node under bad mouthing attack.]{\includegraphics[trim=0 325 650 0,clip, scale = 0.40]{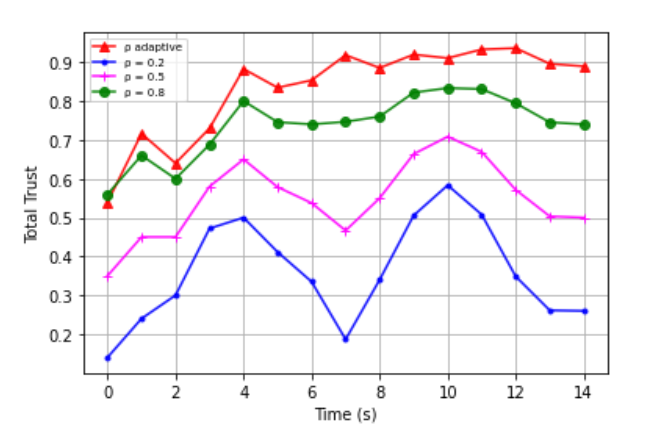}%
\label{bad_mouthing_attack}}
\hfil
\subfloat[Total trust values of a normal node under ballot-stuffing attack.]{\includegraphics[trim=0 325 650 0,clip, scale = 0.40]{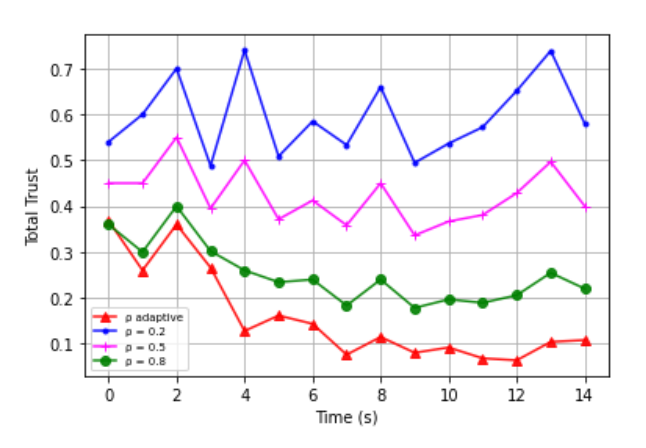}%
\label{ballot_stuffing_attack}}
\caption{Trust variations under bad mouthing and ballot-stuffing attack}
\label{Trust variations under bad mouthing and ballot-stuffing attack}
\end{figure}

The objective of this evaluation is to show the robustness of our trust management system against two trust based attacks, namely Bad-mouthing attack and Ballot-stuffing attack. False recommendations provided by the attackers deteriorates the total trust value of good node. Figure \ref{bad_mouthing_attack} displays our adjustable confidence factor $\rho$ helps the normal node to maintain a high trust value over time compared to the fixed values of $\rho$. Under Ballot-stuffing attack, a good nodes' total trust can be deteriorated because of low recommendations (in the form of indirect trust) it receives from the attackers. Figure \ref{ballot_stuffing_attack} shows that, the adjustable confidence factor $\rho$ facilitates resiliency against Ballot-stuffing attack and helps to keep the trust value of a malicious node low compared to fixed values of $\rho$.

\begin{figure}[!t]
\centering
\includegraphics[trim=3 325 650 4,clip, scale = 0.50]{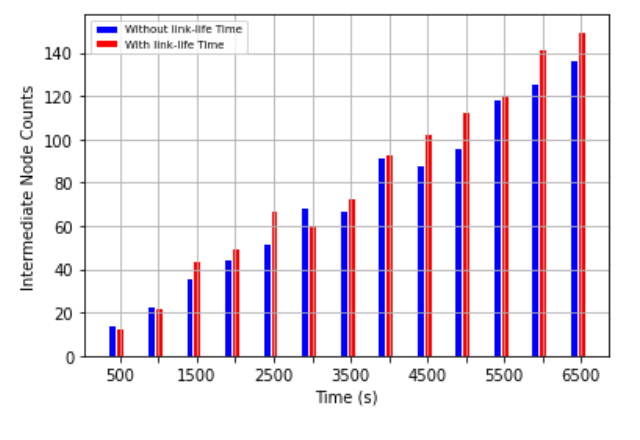}
\caption{Effect of link-life time on neighbour selection}
\label{Effect of Link-life Time on Neighbour Selection}
\end{figure}

\begin{figure*}[!t]
\centering
\subfloat[Response time required by the agent (total nodes = 64)]{\includegraphics[trim=110 265 110 283,clip, scale = 0.3]{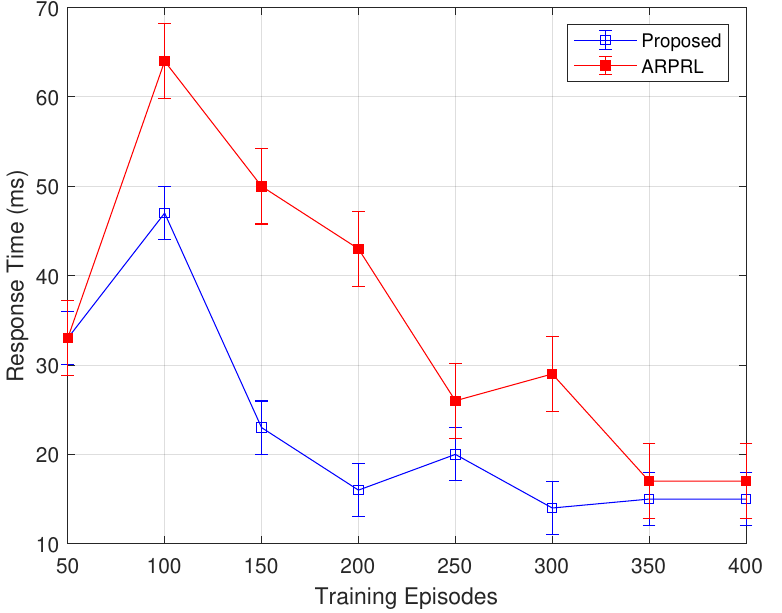}%
\label{computation_time_in_ms}}
\hfil
\subfloat[Effect of topology changes at episode 2001 (total nodes = 64)]{\includegraphics[trim=100 265 115 283,clip, scale = 0.3]{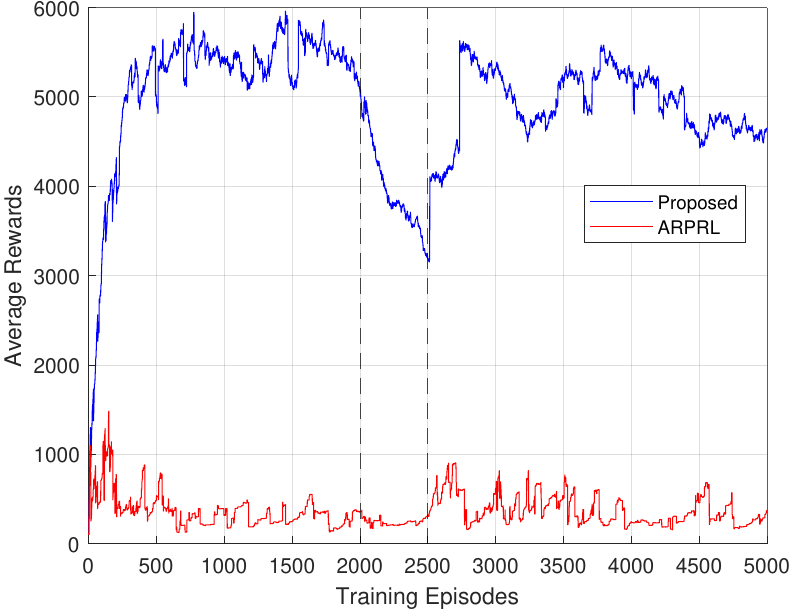}%
\label{rewards_both}}
\hfil
\subfloat[Number of hops required to reach destination]{\includegraphics[trim=110 265 110 283,clip, scale = 0.3]{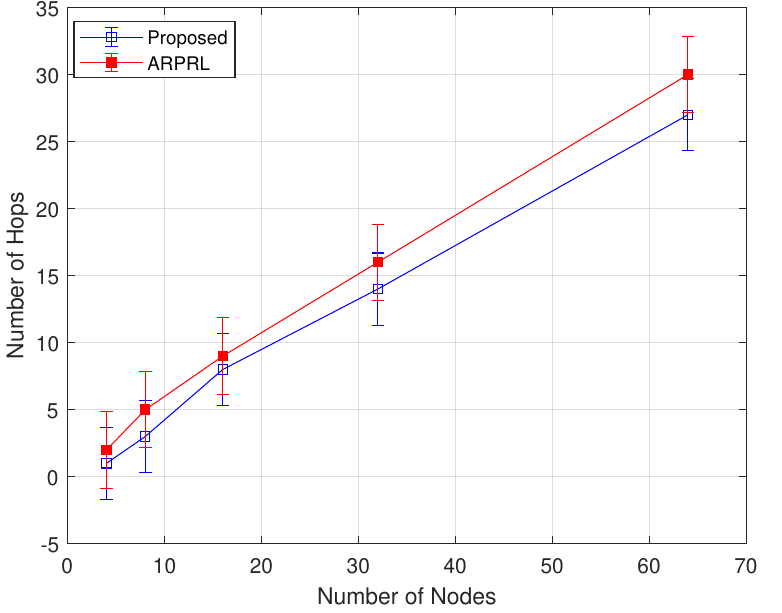}%
\label{hop_count}}
\hfil
\subfloat[Number of packets received with the size of the network]{\includegraphics[trim=100 265 110 283,clip, scale = 0.3]{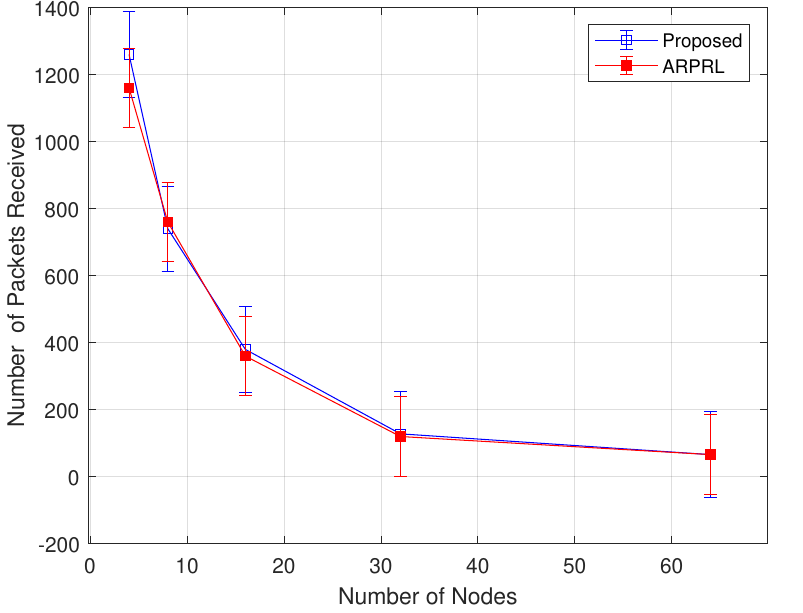}%
\label{packets_received}}
\caption{Performance comparison of proposed framework and ARPRL in terms of response time, adaptability to network changes, average hop counts and number of packets received}
\label{comparison_between_QSAR_ARPRL}
\end{figure*}
\subsection{Effect of Link-life Time}
A neighbour is considered highly stable if its' connection period (link-life time) with the sender is higher compared to other one-hop neighbours. For this obvious reason, the intermediate node count should be higher if link-life time is considered for neighbour selection. The objective of this analysis is to demonstrate if a node was chosen more times when it was a stable neighbour with high link-life time than when it was unstable. To analyse the effect of link-life time, we kept the velocity range of a randomly selected node $n_{i}$ same for a particular simulation period. Then we collected its intermediate node counts by observing how many times other nodes $n_{w}$ (where $w\in N$ and N is the total number of nodes in the network) choose $n_{i}$ as an forwarder for both when link-life time was considered and not considered as a neighbour selection criteria. Figure \ref{Effect of Link-life Time on Neighbour Selection} shows that on average a node receives more requests to forward packets when it stays connected with the senders for a longer period of time.
\subsection{Performance Comparison with ARPRL}
To demonstrate the efficacy of proposed framework in terms of achieving QoS requirements, we compare the performance of proposed framework with a recent routing technique ARPRL \cite{wu2018reinforcement}. The reasons behind choosing ARPRL for comparison are due to the following similarities of ARPRL with proposed framework as ARPRL has applied 1) Q-learning algorithm, 2) link-life time to select neighbours, 3) adjustable learning rate to update the neighbour information. As ARPRL has not considered any security metric, we consider a network of 64 nodes with no attacker for comparison.  
\subsubsection{Response Time}
Response time for any RL based neighbour selection algorithm indicates the time required to learn the topology of the network from scratch for correct decision making. To simulate the response time, we train a source node at different training episodes from 50 to 400. Then, calculate the corresponding response time required by the source to reach the destination (Figure \ref{computation_time_in_ms}) via highly stable intermediate nodes. A training episode starts when a node has a packet to send and ends when the packet reaches to the destination. We observe that, until 300 training episodes, proposed framework exhibits a faster response compared to ARPRL. The fixed and adjustable learning rate and a precise neighbour set adopted in proposed framework let the model learn the routing policies quicker than ARPRL. The reason that the response time of ARPRL is almost close to proposed framework after 350 training episodes is because by that time ARPRL has managed to gather enough information about the topology of the network which helps the protocol to respond as fast as proposed framework. In brief, on average a node (RL agent) in our proposed method responds upto 54\% faster than ARPRL method.
\subsubsection{Adaptability to dynamic network changes}
The adaptability of RL based routing algorithm is defined as the model's ability to learn a new routing policy when the current route becomes unavailable due to a change in network topology. To demonstrate the effect of change in network topology, we first train both proposed framework and ARPRL for 2000 episodes. Until episode 2000, relative velocities of different nodes are kept the same to keep the network topology constant. At episode 2001, we reinitialize the Q-value of a randomly chosen node (which was serving as an intermediate node to the destination) and re-train both models up to episode 5000. Q-value of a randomly chosen intermediate node (that falls under a popular route to destination) is re-initialised to capture the effect of a broken link. We calculate the total reward collected at each episode and to make the comparison acceptable, we consider the same initial reward value for both of the methods. We consider average rewards per episode as a performance metric to demonstrate network adaptability to topology changes. An aggregated high average reward is an outcome of a reasonably good learned policy. Figure \ref{rewards_both} shows that, for proposed framework, the average rewards increases sharply until the learning converges at episode 400. From episode 400 to episode 2000, the average reward remains almost stable. On the other hand, the average reward fluctuates in ARPRL without settling down to a fixed policy. Average rewards in proposed framework and ARPRL drop when the topology is changed at 2001. However, even though the average reward drops more heavily in proposed framework, the reward is still higher than that of ARPRL.
\subsubsection{Average hop counts}
Average hop count defines the total number of hops (steps) required by a packet to reach the destination. The average hop count increases with the size of the network. Less hop counts are desirable to minimise the latency of packet delivery. In our network configuration setting, we consider a minimum of 1 and a maximum of 3 one-hop neighbours for each node. We observe on average a slight improvement to the average hops count (Figure \ref{hop_count}) for large network size.
\subsubsection{Network Packets Received}
We capture the number of packets received for both proposed framework and ARPRL. With the increase in the total number of nodes in the network, the distance between source and destination extends causing the possibility of increased channel collisions and decreased connectivity \cite{wu2018reinforcement}. This results in low packet reception. Despite the other added requirements and improvements of the proposed framework, it does not suffer any performance loss to the number of packets received (Figure \ref{packets_received}).
\subsection{Performance Comparison with ONMC method}
\begin{figure}[t!]
     \centering
     \includegraphics[trim=100 265 110 275,clip, scale = 0.3]{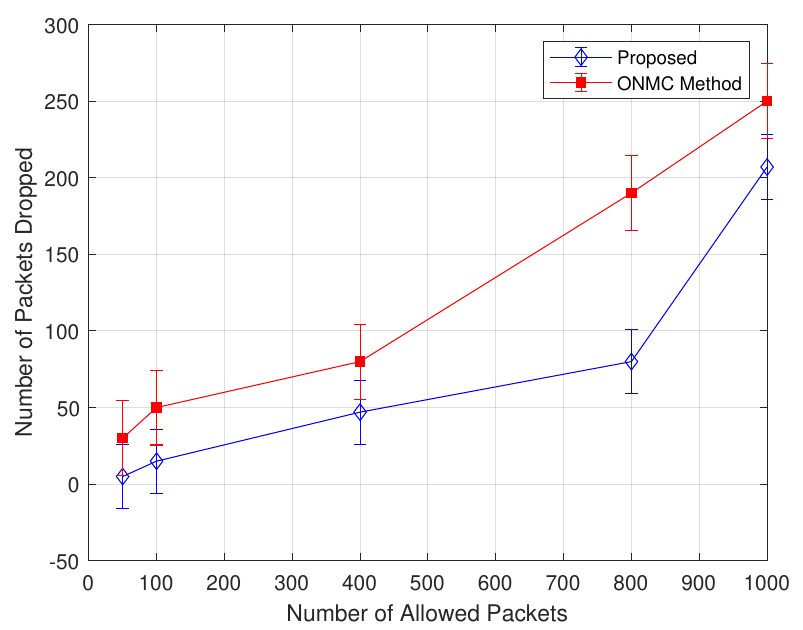}
     \caption{Performance comparison of proposed framework with ONMC method in terms of number of packets dropped}
     \label{fig:comparison_with_ONMC}
 \end{figure}

To show the performance improvement of our proposed framework in the security context, we compare our method with the existing on-policy Monte Carlo (ONMC) learning based secure routing method \cite{usaha2006identifying} in terms of the number of packets dropped. To keep the percentage of attackers ($20\%$) the same, we use the same number of normal and malicious nodes as applied ONMC method. We choose Grayhole attackers as the malicious nodes due to severe security threats they impose on the overall performance of the network compared to Blackhole attackers \cite{mohanapriya2014modified}. Figure \ref{fig:comparison_with_ONMC} shows that our proposed proposed framework reduces the amount of packet dropping up to $57 \%$ compared to ONMC routing method.
\subsection{Overhead Analysis of Proposed Framework}
Execution of proposed framework takes place in two steps: (1) trusted neighbour list $N_{T}$ formation from all the available one-hop neighbours, (2) Q-table generation from $N_{T}$ by calculating the link-life of each neighbour $n$ where $n \in N_{T}$. In worst-case scenario, when all the one-hop neighbours are normal (trustworthy) in a network of $N$ nodes, time complexity of the proposed framework becomes $O(N^{2})$. Table \ref{Overhead_Analysis} shows an overview of space and time complexity information of different RL based routing methods.
\begin{table}[!t]
\scriptsize
\renewcommand{\arraystretch}{1} 
\caption{Comparison of overhead}
\label{Overhead_Analysis}
\centering
\begin{tabular}{c c c c c c c c c c c}
\hline
Method & Stored Information & \begin{tabular}[c]{@{}l@{}}Time\\Complexity \end{tabular}\\
\hline
Q-routing & Q-table &$O(N)$ \\
CQ-routing & Confidence and Q-table & $O(N^{2})$ \\
ONMC Routing & Trust and Q-table &$O(N^{2})$ \\
ARPRL & Q-table &$O(N)$ \\
Proposed & Trust and Q-table &$O(N^{2})$ \\ 
\hline
\end{tabular}
\end{table}
\section{Conclusion and Future Research Directions}
This work presents a Q-learning based framework for intermediate node selection in VANET using trust and link-life time. Our proposed framework can achieve comparable performance in terms of average hop counts, number of received packets at the destination and response time against a baseline method ARPRL. Our proposed trust model considers recent interactions in trust estimation by taking into account the time of observation and can cope with four different types of attackers namely Blackhole, Grayhole, Bad mouthing and Ballot-stuffing attack. We present an adjustable learning rate in the Q-learning algorithm by taking into account the relative velocities of neighbour nodes. We demonstrate that our adjustable learning rate can help network nodes to adapt changes in network topology. Although we validate our results using widely accepted simulators, a potential future work would be to investigate the performance of such framework on real world vehicular mobility traces. Again, the calculation of the learning rate requires the correct selection of the velocity threshold. How to select an optimal velocity threshold that will achieve desirable performance in all possible scenarios is also one of the future research directions.
\section{Acknowledgment}
The work has been supported by the Cyber Security Research Centre Limited whose activities are partially funded by the Australian Government’s Cooperative Research Centres Programme.
\bibliographystyle{ieeetr}

\begin{thebibliography}{10}

\bibitem{sumayya2015vanet}
P.~Sumayya and P.~Shefeena, ``Vanet based vehicle tracking module for safe and efficient road transportation system,'' {\em Procedia Computer Science}, vol.~46, pp.~1173--1180, 2015.

\bibitem{wahid2019holistic}
A.~Wahid, H.~Yasmeen, M.~A. Shah, M.~Alam, and S.~C. Shah, ``Holistic approach for coupling privacy with safety in vanets,'' {\em Computer networks}, vol.~148, pp.~214--230, 2019.

\bibitem{javed2016interrelation}
M.~A. Javed and E.~B. Hamida, ``On the interrelation of security, qos, and safety in cooperative its,'' {\em IEEE Transactions on Intelligent Transportation Systems}, vol.~18, no.~7, pp.~1943--1957, 2016.

\bibitem{malhi2020security}
A.~K. Malhi, S.~Batra, and H.~S. Pannu, ``Security of vehicular ad-hoc networks: A comprehensive survey,'' {\em Computers \& Security}, vol.~89, p.~101664, 2020.

\bibitem{hussain2020trust}
R.~Hussain, J.~Lee, and S.~Zeadally, ``Trust in vanet: A survey of current solutions and future research opportunities,'' {\em IEEE transactions on intelligent transportation systems}, vol.~22, no.~5, pp.~2553--2571, 2020.

\bibitem{mehdi2017game}
M.~M. Mehdi, I.~Raza, and S.~A. Hussain, ``A game theory based trust model for vehicular ad hoc networks (vanets),'' {\em Computer Networks}, vol.~121, pp.~152--172, 2017.

\bibitem{nazib2021reinforcement}
R.~A. Nazib and S.~Moh, ``Reinforcement learning-based routing protocols for vehicular ad hoc networks: A comparative survey,'' {\em IEEE Access}, vol.~9, pp.~27552--27587, 2021.

\bibitem{yau2012reinforcement}
K.-L.~A. Yau, P.~Komisarczuk, and P.~D. Teal, ``Reinforcement learning for context awareness and intelligence in wireless networks: Review, new features and open issues,'' {\em Journal of Network and Computer Applications}, vol.~35, no.~1, pp.~253--267, 2012.

\bibitem{boyan1994packet}
J.~A. Boyan and M.~L. Littman, ``Packet routing in dynamically changing networks: A reinforcement learning approach,'' in {\em Advances in neural information processing systems}, pp.~671--678, 1994.

\bibitem{kumar1999confidence}
S.~Kumar and R.~Miikkulainen, ``Confidence based dual reinforcement q-routing: An adaptive online network routing algorithm,'' in {\em IJCAI}, vol.~99, pp.~758--763, Citeseer, 1999.

\bibitem{wu2018reinforcement}
J.~Wu, M.~Fang, and X.~Li, ``Reinforcement learning based mobility adaptive routing for vehicular ad-hoc networks,'' {\em Wireless Personal Communications}, vol.~101, no.~4, pp.~2143--2171, 2018.

\bibitem{usaha2006identifying}
W.~Usaha and K.~Maneenil, ``Identifying malicious nodes in mobile ad hoc networks using a reputation scheme based on reinforcement learning,'' in {\em TENCON 2006-2006 IEEE Region 10 Conference}, pp.~1--4, IEEE, 2006.

\bibitem{maneenil2005preventing}
K.~Maneenil and W.~Usaha, ``Preventing malicious nodes in ad hoc networks using reinforcement learning,'' in {\em 2005 2nd International Symposium on Wireless Communication Systems}, pp.~289--292, IEEE, 2005.

\bibitem{zhang2018machine}
D.~Zhang, F.~R. Yu, and R.~Yang, ``A machine learning approach for software-defined vehicular ad hoc networks with trust management,'' in {\em 2018 IEEE Global Communications Conference (GLOBECOM)}, pp.~1--6, IEEE, 2018.

\bibitem{zhang2018deep}
D.~Zhang, F.~R. Yu, R.~Yang, and H.~Tang, ``A deep reinforcement learning-based trust management scheme for software-defined vehicular networks,'' in {\em Proceedings of the 8th ACM Symposium on Design and Analysis of Intelligent Vehicular Networks and Applications}, pp.~1--7, 2018.

\bibitem{su2011prevention}
M.-Y. Su, ``Prevention of selective black hole attacks on mobile ad hoc networks through intrusion detection systems,'' {\em Computer Communications}, vol.~34, no.~1, pp.~107--117, 2011.

\bibitem{soleymani2021security}
S.~A. Soleymani, S.~Goudarzi, M.~H. Anisi, M.~Zareei, A.~H. Abdullah, and N.~Kama, ``A security and privacy scheme based on node and message authentication and trust in fog-enabled vanet,'' {\em Vehicular Communications}, vol.~29, p.~100335, 2021.

\bibitem{hussain2019integration}
R.~Hussain, F.~Hussain, and S.~Zeadally, ``Integration of vanet and 5g security: A review of design and implementation issues,'' {\em Future Generation Computer Systems}, vol.~101, pp.~843--864, 2019.

\bibitem{zhou2020distributed}
M.~Zhou, L.~Han, H.~Lu, and C.~Fu, ``Distributed collaborative intrusion detection system for vehicular ad hoc networks based on invariant,'' {\em Computer Networks}, p.~107174, 2020.

\bibitem{littman1993distributed}
M.~Littman and J.~Boyan, ``A distributed reinforcement learning scheme for network routing,'' in {\em Proceedings of the international workshop on applications of neural networks to telecommunications}, pp.~45--51, Erlbaum Hillsdale, NJ, USA, 1993.

\bibitem{nabil2019predicting}
M.~Nabil, A.~Hajami, and A.~Haqiq, ``Predicting the route of the longest lifetime and the data packet delivery time between two vehicles in vanet,'' {\em Mobile Information Systems}, vol.~2019, 2019.

\bibitem{kudva2021scalable}
S.~Kudva, S.~Badsha, S.~Sengupta, H.~La, I.~Khalil, and M.~Atiquzzaman, ``A scalable blockchain based trust management in vanet routing protocol,'' {\em Journal of Parallel and Distributed Computing}, vol.~152, pp.~144--156, 2021.

\bibitem{shabut2014recommendation}
A.~M. Shabut, K.~P. Dahal, S.~K. Bista, and I.~U. Awan, ``Recommendation based trust model with an effective defence scheme for manets,'' {\em IEEE Transactions on mobile computing}, vol.~14, no.~10, pp.~2101--2115, 2014.

\bibitem{pearl2014probabilistic}
J.~Pearl, {\em Probabilistic reasoning in intelligent systems: networks of plausible inference}.
\newblock Elsevier, 2014.

\bibitem{yager1987dempster}
R.~R. Yager, ``On the dempster-shafer framework and new combination rules,'' {\em Information sciences}, vol.~41, no.~2, pp.~93--137, 1987.

\bibitem{shafer1992dempster}
G.~Shafer, ``Dempster-shafer theory,'' {\em Encyclopedia of artificial intelligence}, vol.~1, pp.~330--331, 1992.

\bibitem{wei2014security}
Z.~Wei, H.~Tang, F.~R. Yu, M.~Wang, and P.~Mason, ``Security enhancements for mobile ad hoc networks with trust management using uncertain reasoning,'' {\em IEEE Transactions on Vehicular Technology}, vol.~63, no.~9, pp.~4647--4658, 2014.

\bibitem{hoffman2009survey}
K.~Hoffman, D.~Zage, and C.~Nita-Rotaru, ``A survey of attack and defense techniques for reputation systems,'' {\em ACM Computing Surveys (CSUR)}, vol.~42, no.~1, pp.~1--31, 2009.

\bibitem{zadeh1979validity}
L.~A. Zadeh, {\em On the validity of Dempster's rule of combination of evidence}.
\newblock Electronics Research Laboratory, College of Engineering, University of~…, 1979.

\bibitem{sun2006trust}
Y.~L. Sun, Z.~Han, W.~Yu, and K.~R. Liu, ``A trust evaluation framework in distributed networks: Vulnerability analysis and defense against attacks,'' in {\em Proceedings IEEE INFOCOM 2006. 25TH IEEE International Conference on Computer Communications}, pp.~1--13, IEEE, 2006.

\bibitem{chen2005dempster}
T.~M. Chen and V.~Venkataramanan, ``Dempster-shafer theory for intrusion detection in ad hoc networks,'' {\em IEEE Internet Computing}, vol.~9, no.~6, pp.~35--41, 2005.

\bibitem{ns2}
N.~wiki, ``The network simulator - ns-2,'' May 2016.

\bibitem{mohanapriya2014modified}
M.~Mohanapriya and I.~Krishnamurthi, ``Modified dsr protocol for detection and removal of selective black hole attack in manet,'' {\em Computers \& Electrical Engineering}, vol.~40, no.~2, pp.~530--538, 2014.

\end{thebibliography}

\end{document}